\documentclass[twocolumn,showpacs,preprintnumbers,amsmath,amssymb]{revtex4}

\def\undersim#1{\setbox9\hbox{${#1}$}{#1}\kern-\wd9\lower
    2.5pt \hbox{\lower\dp9\hbox to \wd9{\hss $_\sim$\hss}}}
\usepackage{amssymb}
\usepackage{amsmath}
\usepackage{graphicx}
\usepackage{color}

\def\undersim#1{\setbox9\hbox{${#1}$}{#1}\kern-\wd9\lower
    2.5pt \hbox{\lower\dp9\hbox to \wd9{\hss $_\sim$\hss}}}

\makeatletter

\newcommand{\Rmnum}[1]{\expandafter\@slowromancap\romannumeral #1@}
\makeatother

\begin{document}

\title{Beam energy dependence of cumulants of the net-baryon, net-charge and deuteron multiplicity distributions in Au+Au collisions at $\sqrt{s_{NN}}=3.0-5.0$ GeV}

\author{Yunxiao Ye}
\affiliation{School of Science, Huzhou University, Huzhou 313000, China}
\affiliation{Department of Physics, Zhejiang University, Hangzhou 310027, China}

\author{Yongjia Wang \footnote{Corresponding author: wangyongjia@zjhu.edu.cn}}
\affiliation{School of Science, Huzhou University, Huzhou 313000, China}
\author{Qingfeng Li \footnote{Corresponding author: liqf@zjhu.edu.cn}}
\affiliation{School of Science, Huzhou University, Huzhou 313000, China}
\affiliation{Institute of Modern Physics, Chinese Academy of Sciences, Lanzhou 730000, China}
\author{Dinghui Lu}
\affiliation{Department of Physics, Zhejiang University, Hangzhou 310027, China}
\author{Fuqiang Wang}
\affiliation{School of Science, Huzhou University, Huzhou 313000, China}
\affiliation{Department of Physics and Astronomy, Purdue University, West Lafayette 47907, USA}

\begin{abstract}
Within the ultra-relativistic quantum molecular dynamics (UrQMD) model, in which the Lorentz-covariant treatment of nuclear mean-field potential is considered, the fluctuations of net-baryon, net-charge and deuterons multiplicity distributions in Au+Au head-on collisions at $\sqrt{s_{NN}}=3.0-5.0$ GeV are calculated.
The results show that the nuclear mean-field potential can significantly enhance the magnitude of baryon number fluctuations in narrow rapidity windows, and this enhancement rapidly weakens with increasing beam energy.
However, for proton and net-charge number fluctuations, the mean-field effects are less noticeable than that for baryon number.
In addition, for net-charge number fluctuations, the negative binomial distribution agrees well with the calculated results at mid-pseudorapidity window. Finally, the event-by-event fluctuations of deuteron number in the coalescence production picture are calculated as well, it is found that its cumulant ratios decrease linearly with increasing the average multiplicity of deuterons per event, i.e., increase with increasing beam energy.

\end{abstract}
\maketitle
\section{Introduction}

A fundamental motivation for studying relativistic heavy-ion collisions (HICs) is to investigate the properties of hot and dense nuclear matter which is believed to undergo a phase transition to the quark-gluon plasma (QGP). Lattice Quantum Chromodynamics (QCD) calculations show that the phase transition from the QGP state to hadronic gas is a smooth crossover at zero net-baryon chemical potential $\mu_B$, while QCD-like models predict that the phase transition is first-order at large $\mu_B$.
The possible existence of an end point of the first-order line, called the critical end point, has attracted considerable attentions, see e.g., Refs. \cite{Stephanov:1998dy,Adams:2005dq,BraunMunzinger:2008tz,Fukushima:2010bq,Asakawa:2015ybt,Luo2017} and references therein. To determine the exact location of this critical point, if it exists, is one of the prime goals of the Beam Energy Scan(BES) program at RHIC.
The fluctuations of conserved charges have been proposed as a sensitive observable to explore the critical point \cite{stephanov1999,stephanov2009,stephanov2011,Aggarwal:2010cw,Kumar:2013cqa,Adamczyk:2017iwn,BESII}.
The STAR Collaboration has found that the quartic cumulant ratio, i.e., $\kappa \sigma^2$ of the net-proton number exhibits a non-monotonic energy dependent behavior in the most central Au+Au collisions near $\sqrt{s_{NN}}$=7.7 GeV \cite{STAR-proton,STAR-charge,PHENIX-experiment-net-charge,Xu2014}.
This would be an expected signal near the critical region and a motivation for the BES-II program at RHIC.
In addition, the fixed target experiments from $\sqrt{s_{NN}}=2.7$ to 4.9 GeV have been proposed at the Compressed Baryonic Matter (CBM) detector at the future Facility for Antiproton and Ion Research (FAIR) for complement \cite{FAIR}. The NA61 experiment at CERN-SPS \cite{NA61SHINE}, as well as dedicated future programs at NICA in Russia \cite{NICA}, J-PARC in Japan \cite{JPARC-HI}, HIAF in China \cite{HIAF} may also contribute to the understanding of the QCD phase diagram.

Recent studies suggested that net-proton fluctuations are also influenced by other non-critical effects such as hadron phase \cite{kitazawa2012,Jan2017}, system volume fluctuations \cite{skokov2013,haojie-2016}, baryon clustering \cite{Shuryak:2018lgd}, and global charge conservation \cite{Sakaida:2014pya,he2016,he:2017}.
In order to understand the behavior of fluctuations at beam energy below $\sqrt{s_{NN}} = 7.7$ GeV, in our previous works,
we calculated the cumulant ratios of net-baryon and net-proton numbers in Au+Au collisions at beam energy $E_{lab} =$ 1.23 GeV/nucleon (experiments have been measured by the HADES Collaboration) by the ultra-relativistic quantum molecular dynamics (UrQMD) model, and it is found that both mean-field potential and clustering can influence the magnitude of cumulant ratios \cite{Jan:2018,Ye2018}.
In present work, we use the UrQMD model to illustrate more explicitly the influence of nuclear mean-field potential and clustering on higher cumulant ratios of the net-baryon, net-charge and deuteron multiplicity distributions in head-on Au+Au collisions at $\sqrt{s_{NN}}=3.0 - 5.0$ GeV which correspond to the future CBM experiment.


\section{Brief descriptions of UrQMD model and fluctuation observables}
In the UrQMD model, hadrons obey Hamilton's equations of motion and can be represented by Gaussian wave packets in phase space.
Previous work has pointed out that mean-field potential plays an essential role for describing HICs, especially at low and intermediate energies \cite{Stoecker:1986ci,Bertsch:1988ik}.
The following density- and momentum dependent potential is adopted as the same as that in Refs. \cite{Isse:2005,Li2008}:
\begin{equation}\label{eq2}
U=\alpha (\frac{\rho}{\rho_0})+\beta (\frac{\rho}{\rho_0})^{\gamma}+\sum_{k=1,2} \frac{t^k_{md}}{\rho_0} \int \frac{f({\bf r},\mathbf{p^{\prime }})d\mathbf{p^{\prime }} }{1+[(\mathbf{p}-\mathbf{p^{\prime }})/a^k_{md}]^2}.
\end{equation}

In high energy region, the Lorentz-covariance treatment on the mean-field potential should be considered since Lorentz contraction effect is significant. In this work, the covariant prescription of the mean field used in the relativistic quantum molecular dynamics (RQMD/S) model \cite{Sorge1989,Maruyama1991,Maruyama:1996rn} is adopted.

Moreover, potentials for ``pre-formed"  particles (string fragments)  have been taken into account in the UrQMD model \cite{Li:2008}.
At energies $\sqrt{s}>5$ GeV, color-string excitation and fragmentation dominate the production of hadrons. During the formation time, potentials for ``pre-formed" particles from color fluxtube fragmentation as well as confined particles are considered. Especially for SPS and RHIC energies, it has been found that the ``pre-formed" hadron potentials play an important role at the early stage of HICs.
In this work, a soft equation of state (EoS) with momentum dependence (SM-EoS) is adopted with the parameter set $\alpha=-110.49$ MeV, $\beta=182.014$ MeV, $\gamma=7/6$, $a^1_{md}=0.443986$ GeV$/$c, $t^1_{md}=-261.416$ MeV, $a^2_{md}=0.147995$ GeV$/$c and $t^2_{md}=200$ MeV for Eq. \ref{eq2}.

The cumulants of multiplicity distributions can be defined on event-by-event basis:
\begin{align}
\begin{split}
&C_1=M=\langle N \rangle, \\
&C_2=\sigma^2=\langle (\delta N)^2\rangle, \\
&C_3=S\sigma^3=\langle (\delta N)^3\rangle, \\
&C_4=\kappa \sigma^4=\langle (\delta N)^4\rangle-3\langle (\delta N)^2 \rangle^2.
\end{split}
\end{align}
Here $\delta N=N- \langle N \rangle$, the angular bracket denotes an average from all events, $N$ is the number of particles in a given acceptance window in a single event.
The cumulants are directly related to the susceptibilities with corresponding conserved charges in the grand-canonical ensemble.
The ratios of cumulants are usually constructed to cancel the unknown volume dependence:
\begin{align}
\begin{split}
&C_2/C_1=\sigma^2/M,\\
&C_3/C_1=S \sigma^3/M,\\
&C_3/C_2=S \sigma,\\
&C_4/C_2=\kappa \sigma^2.
\end{split}
\end{align}
Here $M$ is the mean value, $\sigma$ is the standard deviation, $S$ is the skewness, and $\kappa$ is the kurtosis.

According to the Delta-theorem \cite{luo2012}, the statistical error of the cumulants ratios can be approximated as follows:
\begin{align}\label{eq4}
error(C_r/C_2) \propto \sigma^{(r-2)}/ \sqrt{n}.
\end{align}
Here $n$ is the total number of events. In this work, 15 million Au+Au events for each case are calculated. For most observables, the statistical error are small enough (as can be seen in following sections). The statistical error for $C_4/C_2$ is visible. One can estimate from Eq. \ref{eq4} that, to cut the error in half would require four times the number of events.


\section{Numerical results}
\begin{figure*}[t]
	\centering
	\includegraphics[width = 1\linewidth]{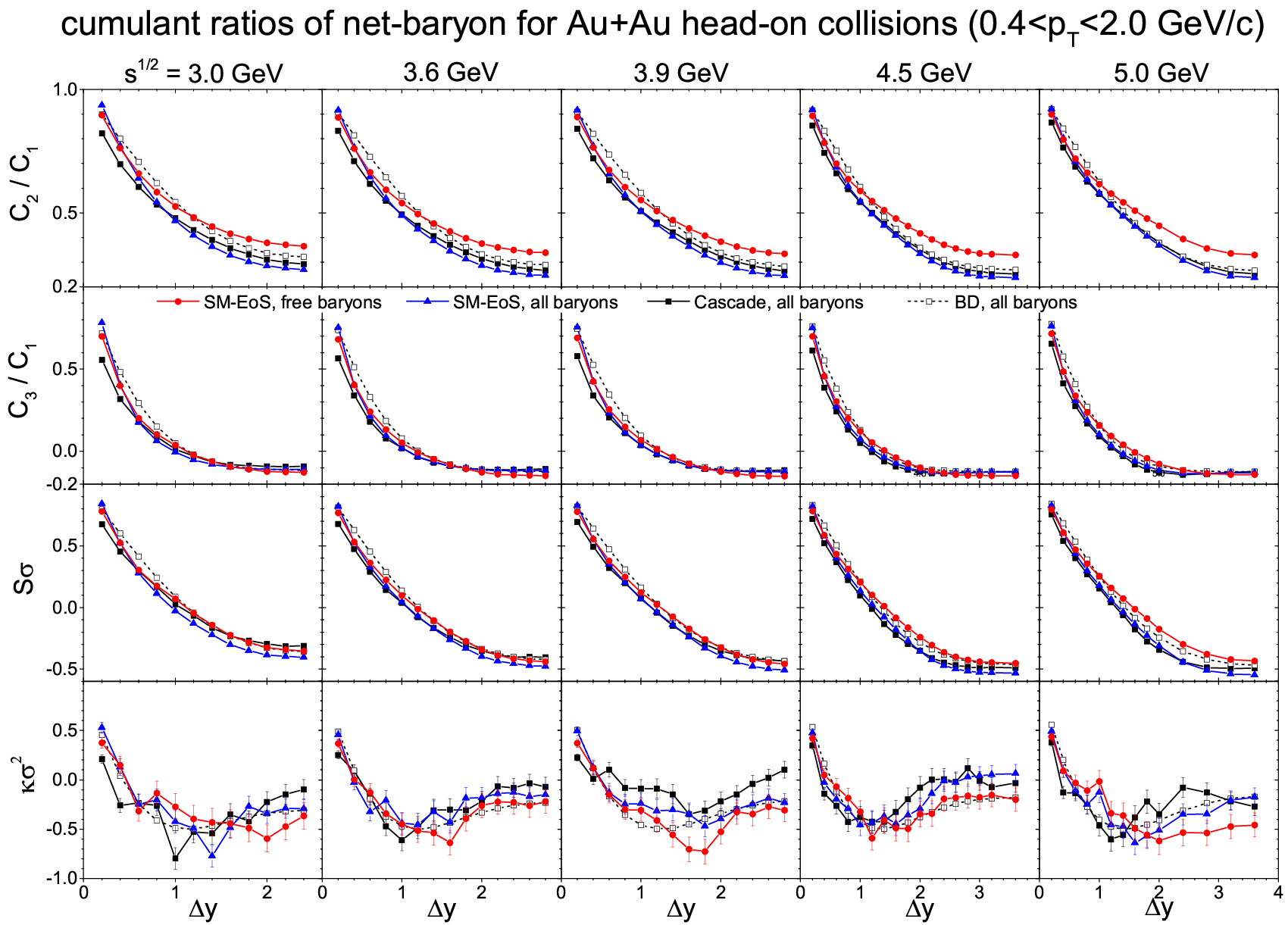}
	\caption{The rapidity window dependence of cumulant ratios for baryon numbers with transverse momentum cut $0.4 <  p_T < 2.0$ GeV$/$c. Black solid squares represent simulation by cascade mode, and open symbols are corresponding binomial distribution baseline (referred as BD. In this case, the average fraction of baryons in the acceptance is obtained from the cascade simulation, while the ratios of cumulants are computed using Eq. \ref{eq6}). Blue triangle symbols represent the result with SM potential. Red circles denote results for free baryon (the baryons that are not bound in fragments). Where error bars are not shown, they fall within the symbols.}
\label{fig1}
\end{figure*}

\subsection{net-baryon}
\begin{figure}[htbp]
	\centering
	\includegraphics[width = 1\linewidth]{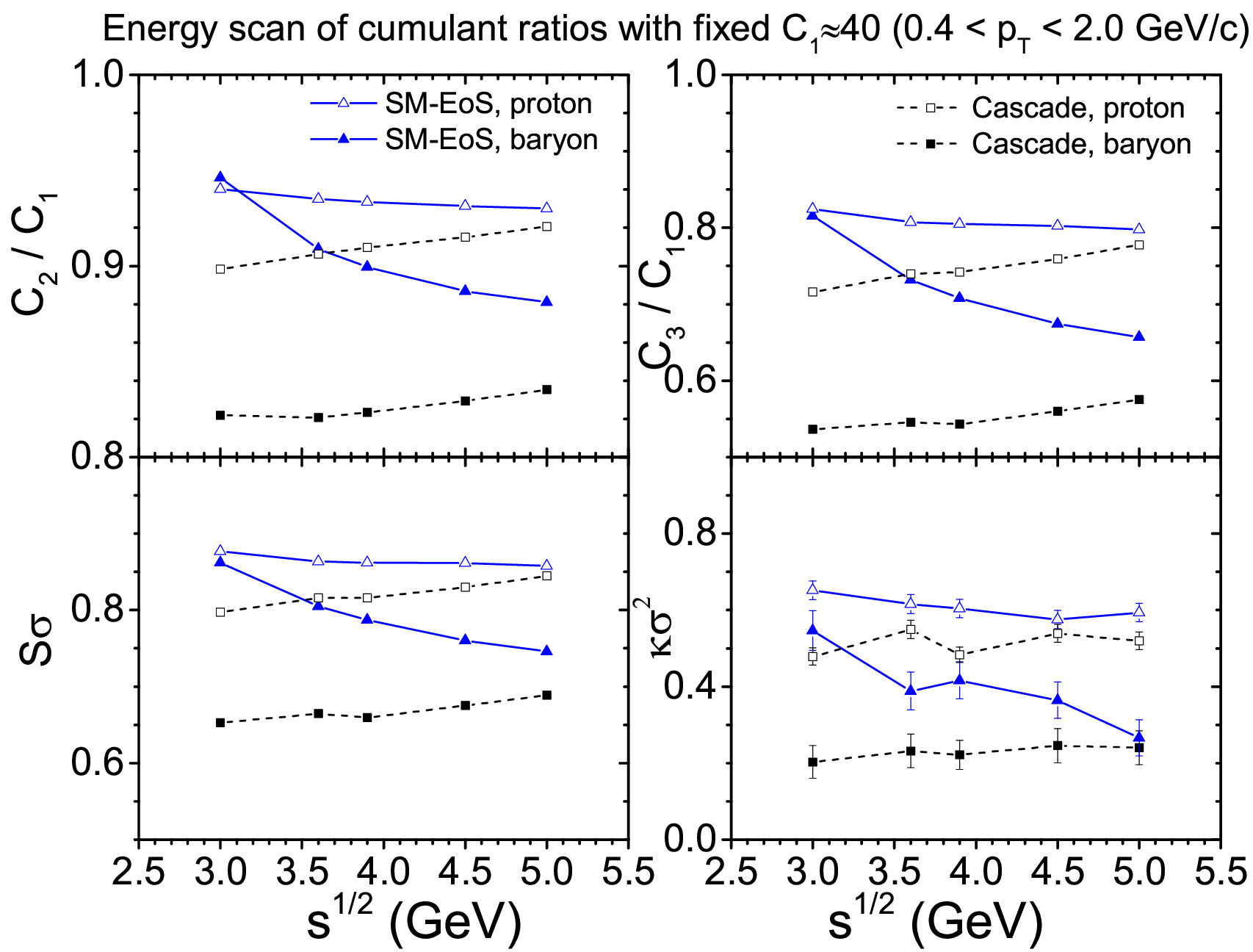}
	\caption{The energy scan of cumulant ratios with fixed baryon number ($C_1 \approx 40$). The results for proton are computed in the same rapidity window as that for baryon. Solid symbols represent the result for baryon number while open symbols for the case of proton number.}
\label{Fig2}
\end{figure}

\begin{figure}[htbp]
\centering
\includegraphics[width = 1\linewidth]{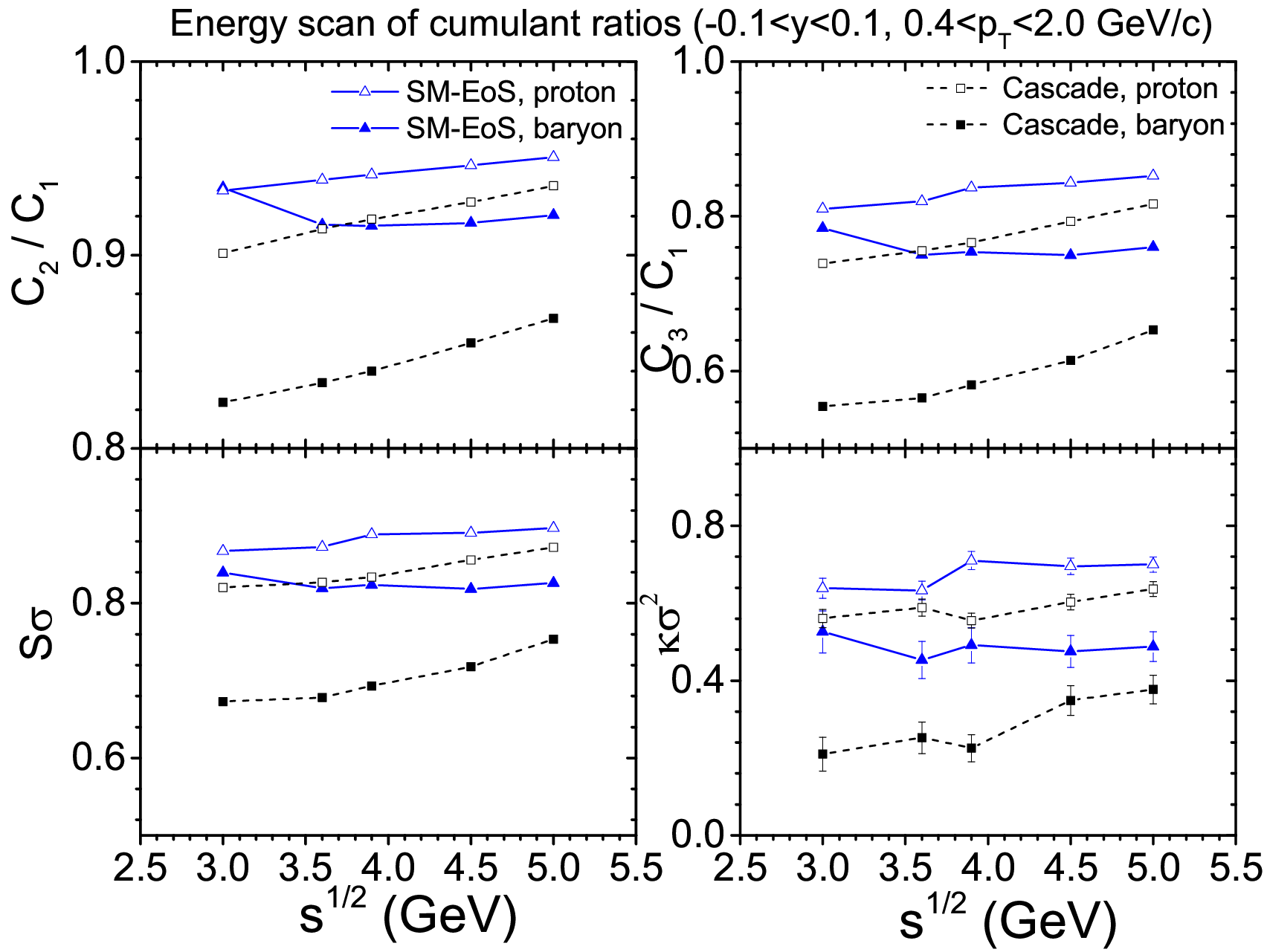}
\caption{The energy scan of cumulant ratios with rapidity window -0.1$<$y$<$0.1. Black lines represent simulation by cascade mode while blue lines represent that with SM-EoS. Solid symbols represent the result for baryon number while open symbols for the case of proton number.}\label{fig22}
\end{figure}

As we know, volume fluctuation is a notorious background and the centrality definition may significantly affect higher moments \cite{westfall2015}. In this work, only head-on collisions (impact parameter $b=0$ fm) are considered. It is presumable that volume fluctuation in collisions with a fixed impact parameter is smaller than that within a range of impact parameter, because the initial colliding geometry in HICs with fixed impact parameter is quite similar. In addition, the formed hot and dense matter in HICs with $b=0$ fm has the largest volume and density, then providing a more clean environment (be less affected by spectators) to study the mean-field effects on the cumulants. As discussed in Ref. \cite{Braun-Munzinger:2016yjz}, the volume (the number of participants) still fluctuates even for a fixed value of impact parameter. Since participants can be precisely counted in transport model, we used similar formulas as the centrality-bin-width correction (CBWC) method \cite{Luo2013,Luo2017} to further reduce the small volume variation caused by different numbers of participants, arising from initial fluctuations even though $b$ is fixed:
\begin{align}
\begin{split}
C_{n} = \frac{\sum_{r} n_{r} C_{n}^{r}}{\sum_{r}n_{r}}.
\end{split}
\end{align}
Here $n_r$ is the number of events with the same number of participants $r$, and $C_{n}^{r}$ is the $n$th-order cumulants of particle multiplicity from those events. We excluded those events where the number of participants is less than 380.
To manifest the effects of the mean-field potential, calculations with pure cascade mode are also presented.
In addition, clustering effect is also taken into account by using an isospin-dependent minimum-span-tree (MST) algorithm to construct clusters \cite{Zhang2012}.
The MST procedure in coordinate and momentum space has been widely used in QMD-like models to recognize fragments in HICs at energies from several tens of MeV/nucleon up to several hundreds of GeV, see e.g., Refs. \cite{Aichelin:1991xy,Li:2016mqd,Sombun:2018yqh}. It has been shown in Ref. \cite{Sombun:2018yqh} that, the deuteron production in p+p, p+A, and A+A collisions at energies from the Facility for Antiproton and Ion Research (FAIR) energy regime up to the CERN Large Hadron Collider energies can be well described with the UrQMD model in combination with the MST algorithm.
In present work, if the relative distance between two protons or two neutrons (neutron and proton) is smaller than 2.8 fm or 3.8 fm, and the relative momentum is smaller than 0.25 GeV/c,
they are considered to belong to the same cluster. In addition, unphysical clusters (such as, di-neutrons or di-protons ) are eliminated by breaking them into free nucleons. We note here that these coalescence parameters used in the present work is consistent with our previous publications, e.g., Refs. \cite{Ye2018,Wang:2014rva,lipc,Wang:2018hsw}. The values of the coalescence parameters are selected empirically to yield a reasonable mass distribution of fragments. Indeed, there is no strict rule for selecting these parameters, usually different parameter sets are used in different models. Fortunately, various dynamical observables, e.g., the collective flow and nuclear stopping power, which have been widely used to deduce the properties of dense nuclear matter is not very sensitive to these empirical parameters within a reasonable range \cite{Wang:2014rva}.
The production of light clusters in HICs from model simulations is an ongoing debate, thermal production and coalescence are two of popular scenarios \cite{Feckova:2016kjx,Oliinychenko:2018ugs,Sombun:2018yqh}.
Different production mechanism may result in different fluctuations \cite{Feckova:2016kjx}. With the MST algorithm to identify deuterons, our calculations partly represent fluctuations in the coalescence production picture.

The rapidity window dependences of the cumulant ratios of net-baryons are shown in Fig.\ref{fig1}.
The interval of transverse momentum $0.4 <  p_T < 2.0$ GeV$/$c is selected by considering the range and capability of detector systems in experiment.
Since the production of anti-baryon can be neglected at this energy scale, we compared the results with BD which represents independent production:
\begin{align}\label{eq6}
\begin{split}
&\sigma^2/M = 1-p, \\
&C_3/C_1 = (1-2p)(1-p), \\
&S\sigma = 1-2p, \\
&\kappa\sigma^2 = 1-6p(1-p). \\
\end{split}
\end{align}
Here $p=C_1/394$ is the average fraction of baryons in a given acceptance. BD is used in the following figures to denote the baseline of the binomial distribution with the $C_1$ obtained from the cascade mode
as the input. We have checked that the baseline of the binomial distribution will not change much if $C_1$ is taken from other scenarios.


In the mid-rapidity region which can be appropriately described by grand-canonical ensemble, the nuclear mean-field potential enhances the magnitude of fluctuations compared to the cascade mode. This is because nuclear potential introduces final multi-particle correlations.
If one compares the results for free baryons with that for all baryons, it is found that the cumulant ratios for free baryons are slightly smaller than that for all baryons at mid-rapidity.
Since at mid-rapidity, free baryons exclude those baryons which are bound in clusters (i.e. more correlations).
With increasing rapidity window, baryon number conservation dominates the results and suppresses the mean field effects, forcing all cumulant ratios to approach to the limit of BD results.
The cumulants of free baryons are less affected by the baryon-number conservation.
Although the tendencies are roughly in accord with the BD baselines, the divergence among calculations and the BD baselines for $\kappa \sigma^2$ seems visible, suggesting that $\kappa \sigma^2$ is more sensitive to the dynamic processes of HICs
and the simple BD cannot fully describe its behavior.

\begin{figure*}[t]
	\centering
	\includegraphics[width = 1\linewidth]{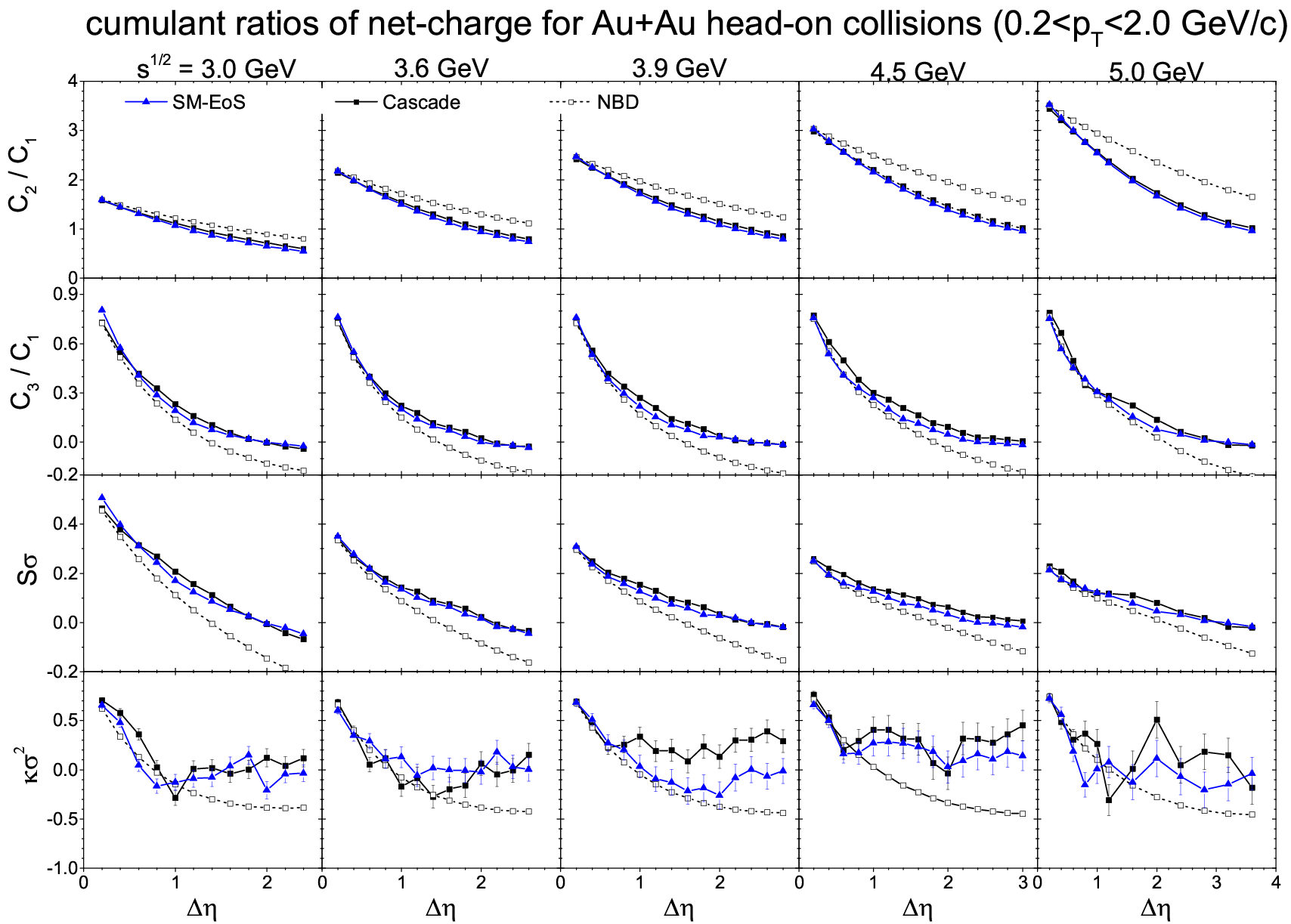}
	\caption{The pseudorapidity window dependence of cumulant ratios for net-charge with transverse momentum cut $0.2<p_T<2.0$ GeV$/$c. Black solid squares represent simulation by cascade mode, and open symbols are corresponding negative binomial distribution baselines. Blue triangles represent the results with SM potential.}
\label{Fig3}
\end{figure*}

Our previous work found that collisions dominate the evolution of cumulants in the compressed stage and the nuclear attractive potential plays a crucial role in the development of fluctuations in the expansion stage \cite{Ye2018}.
With increasing beam energy, baryons move away from each other faster and the effect of mean field weakens rapidly. In Fig. \ref{Fig2}, we carefully chose certain rapidity windows for different cases to fix $C_1 \approx 40$, near $10 \%$ of the total net-baryon number, in order to eliminate the potential misleading information caused by $C_1$ dependence.
The difference between the numerical results of SM-EoS and cascade mode steadily decreases with increasing energy.
This indicates that the correlation derived from mean-field potential is decreasing when the system decouples quickly.

The cumulant ratios in a fixed rapidity window (i.e., -0.1$<$y$<$0.1) are displayed in Fig. \ref{fig22}. The ratios of cumulants for baryon are also decreasing with increasing beam energy in the mean-field case, but the decline is slower than that in Fig. \ref{Fig2} (with fixed $C_1$). Because the decreased $C_1$ weakens the decline of cumulant ratios (e.g., $C_2$/$C_1$). The opposite behavior is observed for the cascade case, the cumulant ratios are increasing with beam energy both with a fixed $C_1$ and with a fixed rapidity window. And this ascent is more evident with fixed rapidity window because of the decreased $C_1$ at high energies. We note here that, the rapidity window (i.e., -0.1$<$y$<$0.1) used in this work is smaller than the commonly employed one (i.e., -0.5$<$y$<$0.5) by STAR Collaboration at energies $\sqrt{s_{NN}}=7.7-200$ GeV. As we have known, with increasing of the collision energy, the rapidity distribution of baryons become more wider, see e.g., Ref. \cite{Yuan:2010ad}. Thus, it is reasonable to apply a smaller rapidity window at lower energies. We have checked that with rapidity window -0.5$<$y$<$0.5, the values of baryon number $C_1$ are 180 and 147 at $\sqrt{s_{NN}}=3.0$ and 5.0 GeV in the cascade case, respectively.
The cumulant ratios obtained with SM-EoS and cascade modes are very close to each other because of the domination of the baryon-number conservation, which also can be seen from Fig. \ref{fig1}.

Cumulant ratios of the net proton number is an experimental proxy observable of net baryon number.
But fluctuations of proton number are more complicated than that of baryon number due to isospin reversal processes in inelastic nucleon-nucleon collisions mediated by the $\Delta$ resonance \cite{kitazawa2012}.
There is no strict conservation law for proton number at every event and this introduces other form of fluctuations.
Especially at intermediate energy, unlike in high energy collisions where nucleons completely lose their original isospin information, some protons retain their initial isospin, so, $N_{proton} \ne \frac{1}{2} N_{baryon}$.
The numerical results indicate that all of these ratios are enhanced by the mean-field potential, but the enhancement is more evident for baryons than for protons. From Figs. \ref{Fig2} and \ref{fig22}, we can infer that, with a small rapidity window around mid-rapidity, the cumulant ratios decrease with collision energy in the presence of mean-field potential, but this trend is being reversed in the cascade case.

\subsection{net-charge}

\begin{figure}[htbp]
	\centering
	\includegraphics[width = 1\linewidth]{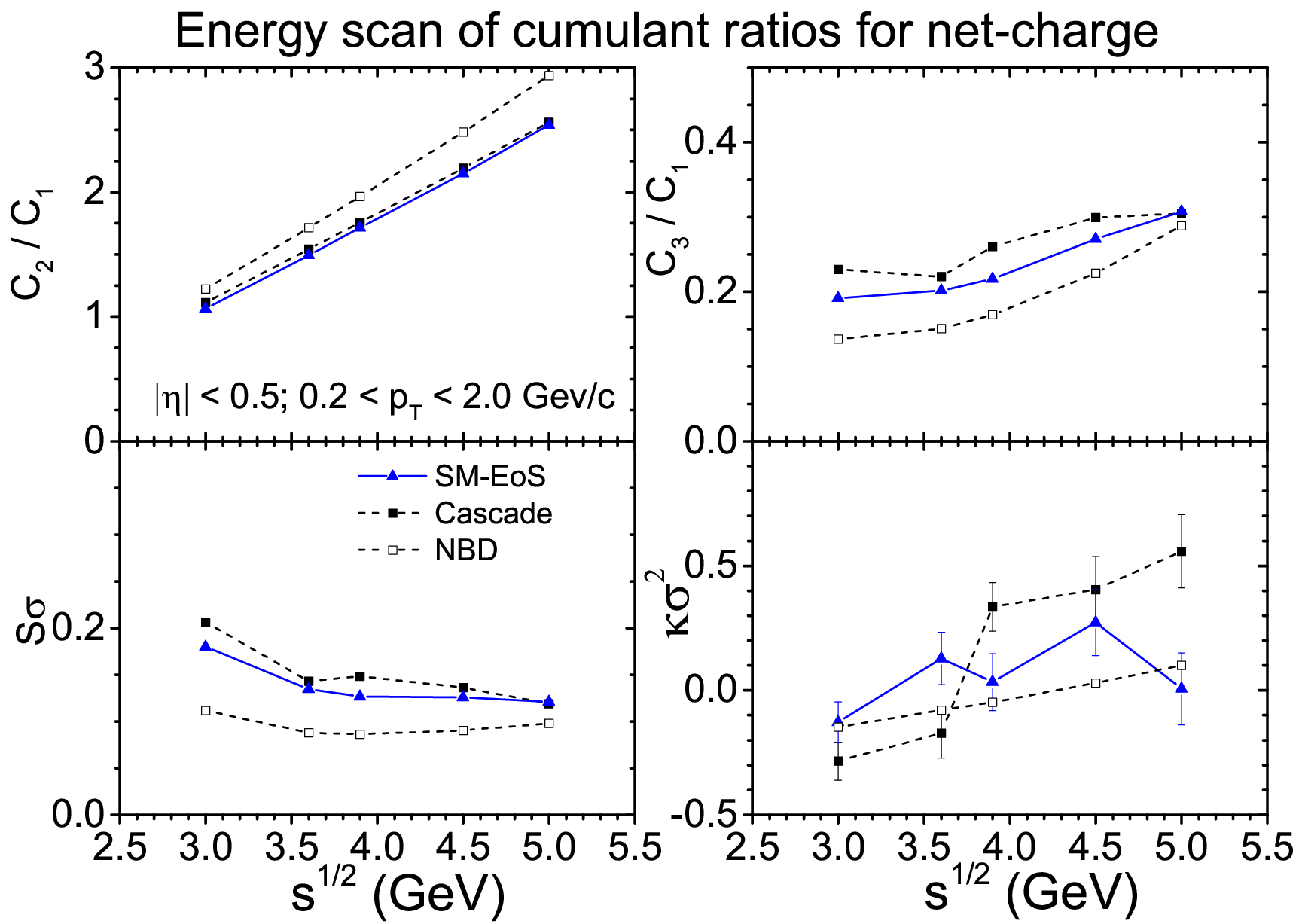}
	\caption{The energy dependence of net-charge fluctuations with fixed acceptance windows: $|\eta|<0.5$, $0.2 < p_T < 2.0$ GeV/c. Notations are the same as in Fig.\ref{Fig3}}
\label{Fig4}
\end{figure}

The pseudorapidity window dependence of cumulant ratios for net-charge ($p_T$ interval 0.2 - 2.0 GeV/c ) are shown in Fig. \ref{Fig3}, and the energy dependence of net-charge fluctuations with fixed pseudorapidity window $|\eta|<0.5$ are show in Fig. \ref{Fig4}. In addition, results from negative binomial distribution (NBD), which has been used to characterize the multiplicity distribution and higher cumulants of charged particle in high-energy \cite{STAR-charge,Westfall2013,Abbott1995}, are also shown.
NBD assumes the production of positive and negative charged particles are unrelated to each other.
A NBD with mean $M$ and variance $\sigma^2$ can be described by two parameters, $q=M/{\sigma^2}$ and $m=Mq/(1-q)$:
\begin{align}
\begin{split}
&C_1 = m(1-q)/q, \\
&C_2 = m(1-q)/q^2, \\
&C_3 = m(q-1)(q-2)/q^3, \\
&C_4 = m(1-q)(6-6q+q^2)/q^4. \\
\end{split}
\end{align}
The above are defined either for positive-charge $C_{n,+}$ and negative-charge $C_{n,-}$ .
The odd-order and even-order cumulant baselines for net-charge can be derived from two negative binomial distribution $C_{n=odd}=C_{n,+}-C_{n,-}$ and $C_{n=even}=C_{n,+}+C_{n,-}$.
NBD is used in the following figures to denote the baseline of the negative binomial distribution with the $C_1$ and $C_2$ calculated by cascade mode as the inputs. For a negative binomial distribution, the probability $q$ should not exceed the value of 1, i.e., $M < \sigma^2$. In general, the results of NBD are in accord with the UrQMD simulations, especially for the energy-dependent behavior of $C_2 / C_1$ and $S \sigma$ (ratios of cumulants with different odd-even parities) at  mid-pseudorapidity.
With increasing pseudorapidity window, the deviation among NBD and UrQMD calculations steadily grows, as the condition $M < \sigma^2$ for NBD is manifestly violated.
In large pseudorapidity window, the production of positive and negative charged particles are always correlated because the total charge is strictly conserved.

Unlike to the case for baryons, the cumulant ratios of net-charge are only sightly influenced by nuclear mean-field potential.
Because with increasing beam energy, more and more charged mesons are produced through inelastic collisions or decay of resonances, meson-baryon scattering may play a dominant role in net-charge fluctuations.

\subsection{deuteron}

It is suggested that a double-peak structure of density fluctuation can be observed from spinodal instability region to critical region with increasing beam energy \cite{jan2012,herold2014,li2017}.
Directly extracting the information of density fluctuations in heavy-ion collisions seems impossible since only the momenta of final state particles are measured in experiment.
However, light fragments (e.g., deuteron, triton) may sensitive to density fluctuations at freeze-out.
Moreover, light nuclei production has been used as a probe of the QCD phase diagram \cite{sun2018}, thus it is worth to study moments of deuteron distribution.

\begin{figure}[htbp]
	\centering
	\includegraphics[width = 1\linewidth]{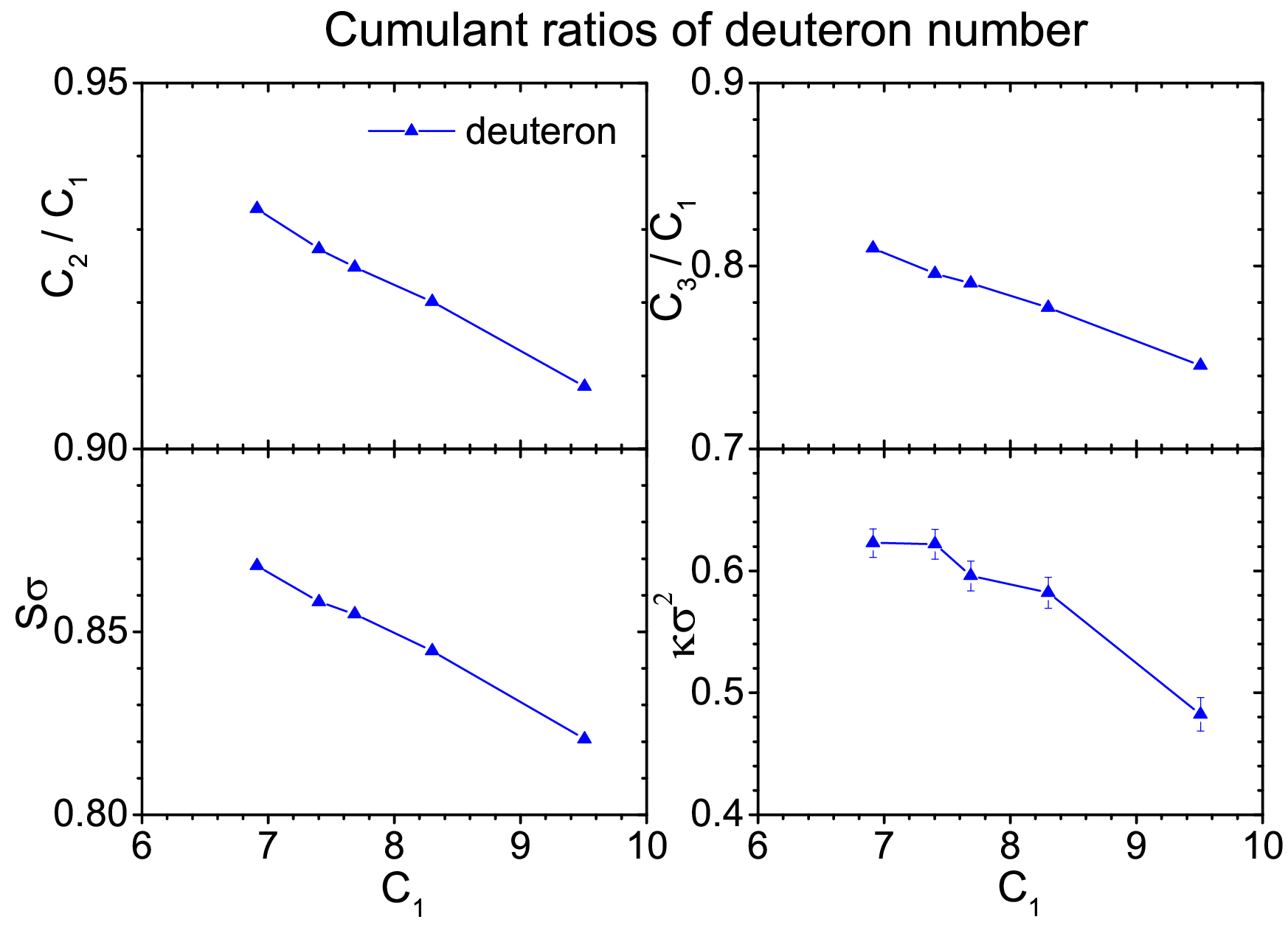}
	\caption{The cumulant ratios as a function of the averaged deuteron multiplicity per event at different collision energies. No acceptance cuts are applied. More deuterons (larger $C_1$) correspond to lower beam energy.}
\label{Fig5}
\end{figure}

Fig.\ref{Fig5} displays the result of energy scan of deuteron cumulant ratios as a function of $C_1$. All cumulants ratios show a monotonous trend and a fairly good linearity with the averaged deuteron multiplicity. It illustrates deuteron number fluctuation tends to monotonically increase with increasing beam energy in the coalescence production picture.


\section{Summary}
We have calculated the cumulant ratios for the net-baryon, net-charge and deuteron multiplicity distributions in Au+Au collisions at $\sqrt{s_{NN}}$=3.0, 3.6, 3.9, 4.5, 5.0 GeV with the UrQMD model, in which the Lorentz-covariance treatment of nuclear mean-field potential is further considered. The results simulated with mean-field potential (SM-EoS) mode were compared to those with cascade mode.
It is found that the nuclear mean-field potential can enhance the magnitude of cumulant ratios of net-baryon number, and that this enhancement decreases with increasing beam energy.
This result implies that the correlation of final baryons established by mean-field potential decreases with increasing collision energy. With a small rapidity window around mid-rapidity, the cumulant ratios decrease with collision energy in the presence of mean-field potential, but this trend is being reversed in the cascade case. The mean-field effects on proton number cumulant ratios is less conspicuous.
On the other hand, it is found that the mean field effects on net-charge fluctuation are less significant and negative binomial distribution can reasonably describe the behavior of cumulant ratios at mid-pseudorapidity. Finally, we calculated the deuteron number fluctuations. The results show that deuteron cumulant ratios are approximately linear in $C_1$ and that fluctuations of deuteron increase with increasing beam energy in the coalescence production picture.

In future study, to obtain an in-depth understanding of the influence of mean-field potential on fluctuations, we plan to investigate the centrality dependence of the cumulant ratios and the volume fluctuations with mean-field mode.
\section{Acknowledgments}
Fruitful discussions with Dr. Jan Steinheimer and Prof. Horst Stoecker are greatly appreciated. The authors acknowledge support of the computing server C3S2 at the Huzhou University. This work is supported in part by the National Natural
Science Foundation of China (Grants No. 11875125, No.11847315, No. 11947410, and  No. 11505057) and the Zhejiang Provincial Natural Science Foundation of China (Grants No. LY19A050005, No. LY18A050002, and No. LY19A050001), and the ``Ten Thousand Talent Program" of Zhejiang province.


\end{document}